\documentclass[11pt]{article}
\usepackage{graphicx}
\usepackage{amssymb,amsmath}
\usepackage[sort&compress,numbers]{natbib}
\usepackage{xcolor}

\begin{document}

\title{
\textbf{Corrections to finite--size scaling in the 3D Ising model
based on non--perturbative approaches \\ and Monte Carlo simulations}
}

\author{J. Kaupu\v{z}s$^{1,2}$ 
\thanks{E--mail: \texttt{kaupuzs@latnet.lv}} \hspace{1ex}, 
R. V. N. Melnik$^3$, 
J. Rim\v{s}\=ans$^{1,2,3}$ \\
$^1$Institute of Mathematics and Computer Science, University of Latvia\\
29 Rai\c{n}a Boulevard, LV--1459 Riga, Latvia \\
$^2$ Institute of Mathematical Sciences and Information Technologies, \\
University of Liepaja, 14 Liela Street, Liepaja LV--3401, Latvia \\
$^3$ The MS2 Discovery Interdisciplinary Research Institute, \\
Wilfrid Laurier University, Waterloo, Ontario, Canada, N2L 3C5}

\maketitle

\begin{abstract}
Corrections to scaling in the 3D Ising model are studied
based on non--perturbative analytical arguments and Monte Carlo (MC) simulation data for
different lattice sizes $L$. 
Analytical arguments show the existence of corrections
with the exponent $(\gamma-1)/\nu \approx 0.38$,
the leading correction--to--scaling exponent being $\omega \le (\gamma-1)/\nu$. 
A numerical estimation of $\omega$ from the susceptibility data 
within $40 \le L \le 2048$ yields $\omega=0.25(33)$. It is consistent
with the statement $\omega \le (\gamma-1)/\nu$, as well as with the value $\omega = 1/8$ of 
the GFD theory. We reconsider the MC estimation of $\omega$ from
smaller lattice sizes to show that it does not lead to conclusive results,
since the obtained values of $\omega$ depend on the particular method chosen.
In particular, estimates ranging from $\omega =1.274(72)$ to $\omega=0.18(37)$
are obtained by four different finite--size scaling methods, using MC data for
thermodynamic average quantities, as well as for partition function zeros.
We discuss the influence of $\omega$ on the estimation of exponents $\eta$ and $\nu$.  
\end{abstract}


\section{Introduction}
\label{intro}

The critical exponents of the three--dimensional (3D) Ising universality class have
been a subject of extensive analytical as well as Monte Carlo (MC) studies during many years.
The results of the standard perturbative renormalization group (RG) methods are well 
known~\cite{Amit,Ma,Justin,Kleinert,PV}. An alternative analytical approach has been proposed 
in~\cite{K_Ann01} and further analyzed in~\cite{K2012}, where this approach is called
the GFD (Grouping of Feynman Diagrams) theory. A review of MC work till 2001 is provided 
in~\cite{HasRev}. More recent papers are~\cite{Has1,Has2,GKR11,KMR_2011,KMR_2013}.

In this paper we will focus on the exponent $\omega$, which describes the leading
corrections to scaling. A particular interest in this subject is caused by recent
challenging non-perturbative results reported in~\cite{K2012}, showing that 
$\omega \le (\gamma-1)/\nu$ holds in the $\varphi^4$ model 
based on a rigorous proof of certain theorem. The scalar 3D $\varphi^4$ model
belongs to the 3D Ising universality class with $(\gamma-1)/\nu \approx 0.38$. 
Therefore, $\omega$ is expected to be essentially smaller than 
the values of about $0.8$ predicted by standard perturbative methods and currently available MC estimations.
The results in~\cite{K2012} are fully consistent with the predictions of the alternative
theoretical approach of~\cite{K_Ann01}, from which $\omega=1/8$ is expected.
We have performed a Monte Carlo analysis of the standard 3D Ising model, using our
data for very large lattice sizes $L$ up to $L=2048$, to clarify whether $\omega$,
extracted from such data, can be consistent with the results of~\cite{K2012} 
and~\cite{K_Ann01}. Since our analysis supports this possibility, we have
further addressed a related question how a decrease in $\omega$ influences the MC estimation of critical
exponents $\eta$ and $\nu$. We have also tested different finite--size scaling methods of estimation $\omega$
from smaller lattice sizes to check whether such methods always give $\omega$ consistent
with $0.832(6)$, as one can be expected from the references in~\cite{Has1}.

Models with the so-called improved Hamiltonians
are often considered instead of the standard Ising model for a better estimation of the
critical exponents~\cite{Has1,Has2}. The basic idea of this approach is to find such Hamiltonian
parameters, for which the leading correction to scaling vanishes.
However, this correction term has to be large enough and
well detectable for the estimation of $\omega$. So, this idea is not very useful in our case.

\section{Analytical arguments}
\label{sec:analytical}

In~\cite{K2012}, the $\varphi^4$ model in the thermodynamic limit has been considered, for which 
the leading singular part of specific heat $C_V^{sing}$ can be expressed as
\begin{equation}
C_V^{sing} \propto \xi^{1/\nu} \left( \int_{k<\Lambda'} [G({\bf k})- G^*({\bf k})] d {\bf k} \right)^{sing} \;,
\label{eq:CVsing}
\end{equation}
assuming the power--law singularity $\xi \sim t^{-\nu}$ of the correlation length $\xi$
at small reduced temperature $t \to 0$. Here
$G({\bf k})$ is the Fourier--transformed two--point correlation function, and $G^*({\bf k})$ is its value
at the critical point.
This expression is valid for any positive $\Lambda' < \Lambda$, where $\Lambda$ is the upper cut-off parameter
of the model, since the leading singularity
is provided by small wave vectors with the magnitude $k = \mid {\bf k} \mid \to 0$ and not by the region 
$\Lambda' \le k \le \Lambda$. In other words, $C_V^{sing}$ is independent of the constant $\Lambda'$.

The leading singularity of specific heat in the form of $C_V^{sing} \propto (\ln \xi)^{\lambda} \xi^{\alpha/\nu}$ 
and the two--point correlation function in the asymptotic form of 
$G({\bf k}) = \sum_{\ell \ge 0} \xi^{(\gamma - \theta_{\ell})/\nu} g_{\ell}(k \xi)$, 
$G^*({\bf k}) = \sum_{\ell \ge 0} b_{\ell} k^{(-\gamma + \theta_{\ell})/\nu}$ with $\theta_0=0$ and $\theta_{\ell}>0$ for $\ell \ge 1$
have been considered in~\cite{K2012}. These expressions are consistent with the conventional scaling hypothesis, $g_{\ell}(k \xi)$
being the scaling functions. The exponent $\lambda$ is responsible for possible logarithmic correction in specific heat,
whereas the usual power--law singularity is recovered at $\lambda=0$.

According to the theorem proven in~\cite{K2012}, the two--point correlation function of the $\varphi^4$ model 
contains a correction with the exponent $\theta_{\ell} = \gamma + 1 -\alpha - d \nu$, if  $C_V^{sing}$ can be calculated
from~(\ref{eq:CVsing}), applying the considered here scaling forms, if the result is $\Lambda'$--independent, and if
the condition $\gamma + 1 -\alpha - d \nu >0$ is satisfied for the critical exponents. Applying the known
hyperscaling hypothesis $\alpha + d \nu = 2$, it yields $\theta_{\ell} = \gamma -1$ for $\gamma >1$.
Apparently, the listed here conditions of the theorem are satisfied for the scalar 3D $\varphi^4$ model.
Since the critical singularities are provided by long--wave fluctuations, the condition of $\Lambda'$--independence is
generally meaningful. The assumption $\xi \sim t^{-\nu}$ (with no logarithmic correction) and the considered here scaling 
forms (with $\lambda =0$), as well as the relation $\gamma + 1 -\alpha - d \nu >0$ (or $\gamma >1$ according
to the hyperscaling hypothesis) are correct for the scalar 3D $\varphi^4$ model, according to the
current knowledge about the critical phenomena. 

The correction with the exponent $\theta_{\ell}$ corresponds to the one with $\omega_{\ell} =\theta_{\ell}/\nu$
in the finite--size scaling. In such a way, the discussed here analytical arguments predict the existence
of a finite--size correction with the exponent $(\gamma-1)/\nu$ in the scalar 3D $\varphi^4$ model.
As discussed in~\cite{K2012}, nontrivial corrections tend to be cancelled in the 2D Ising model, in such a way that
only trivial ones with integer $\theta_{\ell}$ are usually observed.
However, there is no reason to assume such a scenario in the 3D case.
Therefore, the existence of corrections with the exponent $(\gamma-1)/\nu$ is expected
in the 3D Ising model, since it belongs to the same universality class as the 3D $\varphi^4$ model. 
Because this correction is not
necessarily the leading one, the prediction is $\omega \le \omega_{\mathrm{max}}$, where $\omega_{\mathrm{max}} =(\gamma-1)/\nu$
is the upper bond for the leading correction--to--scaling exponent $\omega$.
Using the widely accepted estimates $\gamma \approx 1.24$ and $\nu \approx 0.63$~\cite{Justin} for the 3D Ising model,
we obtain $\omega_{\mathrm{max}} \approx 0.38$. The prediction of the GFD theory~\cite{K_Ann01} is
$\gamma=5/4$, $\nu =2/3$ and, therefore, $\omega_{\mathrm{max}} = 0.375$. Thus, we can state that in any case
 $\omega_{\mathrm{max}}$ is about $0.38$. The value of $\omega$ is expected to be $1/8$ according to
 the GFD theory considered in~\cite{K_Ann01,K2012}.

\section{MC estimation of $\omega$ from finite--size scaling}

\subsection{The case of very large lattice sizes $L \le 2048$}
\label{sec:large}

We have simulated the 3D Ising model on simple cubic lattice with periodic boundary conditions.
The Hamiltonian $H$ of the model is given by
\begin{equation} 
H/T = - \beta \sum_{\langle ij \rangle} \sigma_i \sigma_j \;,
\end{equation}
where $T$ is the temperature measured in energy units, $\beta$ is the coupling constant
and $\langle ij \rangle$ denotes the pairs of
neighboring spins $\sigma_i = \pm 1$.
The MC simulations have been performed with the Wolff
single cluster algorithm~\cite{Wolff}, using its parallel implementation
described in~\cite{KMR_2010}. An iterative method, introduced in~\cite{KMR_2010}, has been
used here to find pseudocritical couplings $\widetilde{\beta}_c(L)$ corresponding
to certain value $U=1.6$ of the ratio $U=\langle m^4 \rangle / \langle m^2 \rangle^2$,
where $m$ is the magnetization per spin. We have evaluated by this method the susceptibility
$\chi = L^3 \langle m^2 \rangle$ an the derivative $\partial Q /\partial \beta$ at 
$\beta = \widetilde{\beta}_c(L)$, where $Q=1/U$. The results for $16 \le L \le 1536$
are already reported in Tab.~1 of our earlier paper~\cite{KMR_2011}. We have
extended the simulations to lattice sizes $L=1728$ and $L=2048$, using approximately
the same number of MC sweeps as for $L=1536$ in~\cite{KMR_2011}.
Thus, Tab.~1 of~\cite{KMR_2011} can be now completed with the new results presented in
Tab.~\ref{tab1} here.

\begin{table}
\caption{The values of $\widetilde \beta_c$, as well as $\chi/L^2$, and
$10^{-3} \partial Q /\partial \beta$ at $\beta = \widetilde{\beta}_c$ depending on $L$.}
\label{tab1}
\begin{center}
\begin{tabular}{|c|c|c|c|}
\hline
\rule[-2mm]{0mm}{7mm}
L & $\widetilde \beta_c$ & $\chi/L^2$  & $10^{-3} \partial Q /\partial \beta$  \\
\hline
2048 & 0.2216546252(66) & 1.1741(27) & 151.1(1.1)  \\
1728 & 0.2216546269(94) & 1.1882(20) & 116.98(87)  \\
\hline
\end{tabular}
\end{center}
\end{table}

\begin{figure}
\begin{center}
\includegraphics[width=0.485\textwidth]{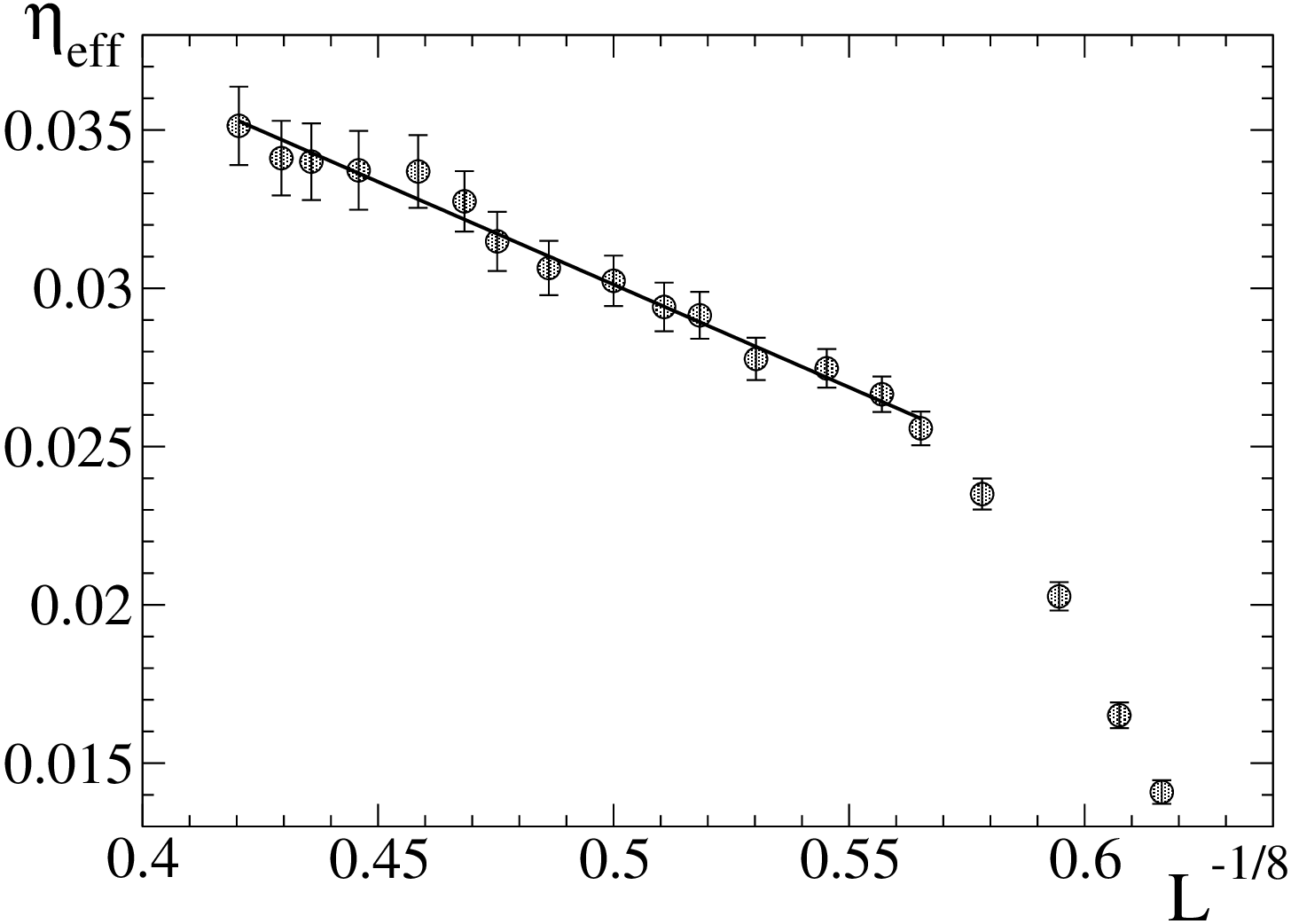}
\hfill
\includegraphics[width=0.485\textwidth]{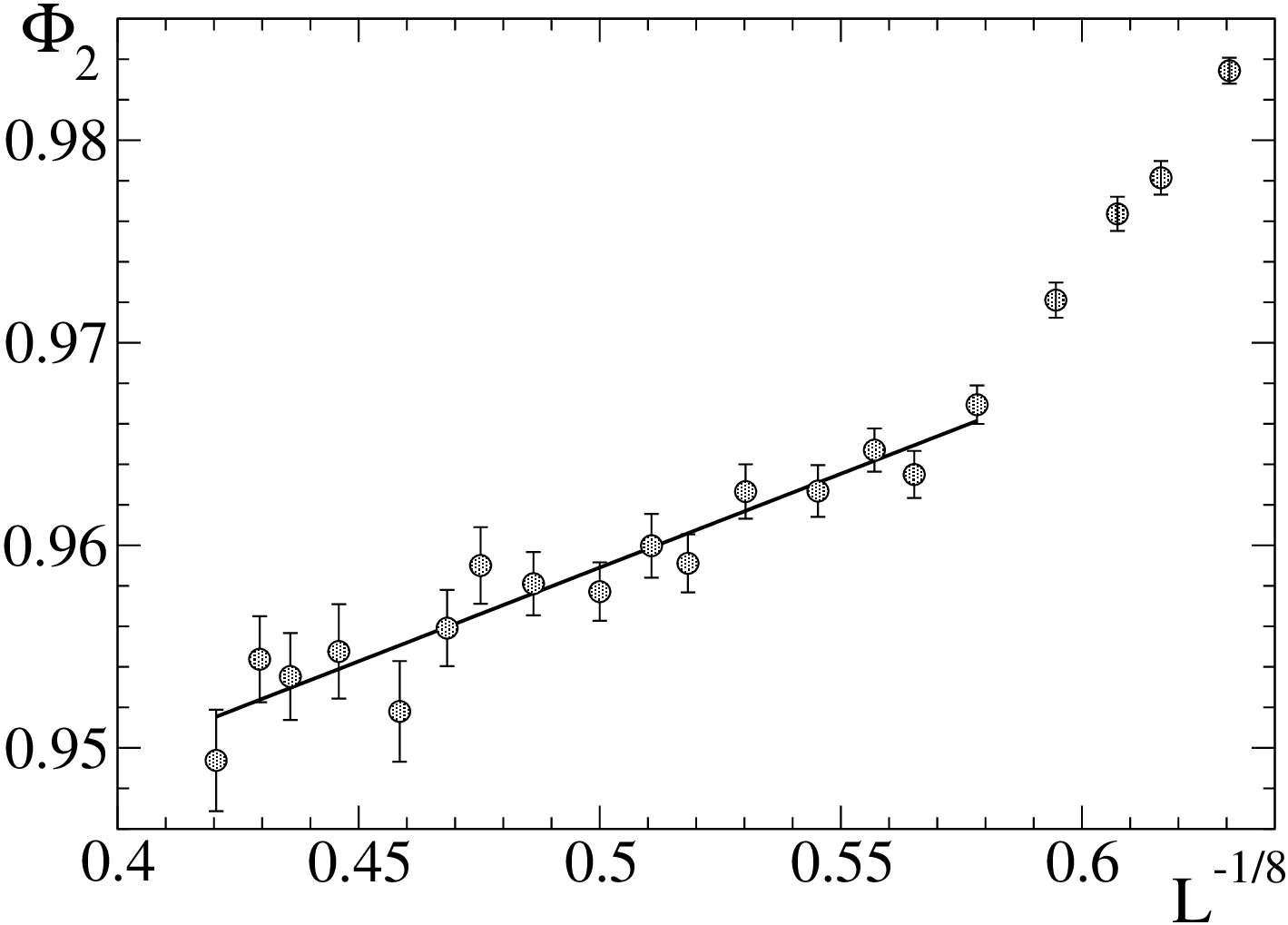}
\end{center}
\caption{The  $\eta_{\mathrm{eff}}$ vs $L^{-1/8}$ (left) and the $\Phi_2(L)$ 
vs $L^{-1/8}$ (right) plots. Straight lines show the linear fits for large
enough lattice sizes $L$.}
\label{fig1}
\end{figure}

The exponent $\omega$ describes corrections to the asymptotic finite--size scaling.
In particular, for the susceptibility at $\beta = \widetilde{\beta}_c(L)$ we have
\begin{equation}
 \chi \propto L^{2-\eta} \left( 1 + a L^{-\omega} + o \left( L^{-\omega} \right) \right) \;.
\label{eq:chi}
\end{equation}
We define the effective exponent $\eta_{\mathrm{eff}}(L)$ as the mean slope of the 
$-\ln \chi$ vs $\ln L$ plot, evaluated by fitting the data within $[L/2,2L]$. 
It behaves asymptotically as $\eta_{\mathrm{eff}}(L)=\eta + \mathcal{O} \left( L^{-\omega} \right)$. 
It has been mentioned in~\cite{KMR_2011} that $\omega$ might be as small as $1/8$, since
the plot of the effective exponent $\eta_{\mathrm{eff}}$ vs $L^{-1/8}$ looks rather linear
for large lattice sizes (see Fig.~6 in~\cite{KMR_2011}).
This observation is confirmed also by the extended here data, as it can be seen from
Fig.~\ref{fig1} (left). 

An estimate of $\omega$ can be obtained by fitting the $\eta_{\mathrm{eff}}(L)$ data.
Here we use a more direct method, which gives similar, but slightly more accurate results.
We consider the ratio $\Phi_b(L) = b^{-4} \chi(bL)/\chi(L/b)$ at $\beta = \widetilde{\beta}_c(L)$, where $b$ is a constant. 
According to~(\ref{eq:chi}), $\Phi_b(L)$ behaves as
\begin{equation}
 \Phi_b(L) = A + B L^{-\omega}
 \label{eq:phi}
\end{equation}
at $L \to \infty$, where $A = b^{-2 \eta}$ and $B = a b^{-2 \eta} \left( b^{-\omega} -b^{\omega} \right)$.
The correction amplitude $B$ is larger for a larger $b$ value, whereas a smaller $b$ value allows us to
obtain more data points for $\Phi_b(L)$. The actual choice $b=2$ is found to be optimal for our data.  
Like the $\eta_{\mathrm{eff}}(L)$ vs $L^{-1/8}$ plot, also the $\Phi_2(L)$ 
vs $L^{-1/8}$ plot can be well approximated by a straight line for large enough lattice sizes, as shown in Fig.~\ref{fig1}.
Thus, $\omega$ could be as small as $1/8$.

\begin{table}
\caption{The values of $\omega$, extracted from the fits of $\Phi_2(L)$ 
to~(\ref{eq:phi}) within $L \in [L_{\mathrm{min}},1024]$, together with
the values of $\chi^2/\mathrm{d.o.f.}$ of the fits.}
\label{tab2}
\begin{center}
\begin{tabular}{|c|c|c|}
\hline
\rule[-2mm]{0mm}{7mm}
$L_{\mathrm{min}}$ & $\omega$  & $\chi^2/\mathrm{d.o.f.}$ \\
\hline
32 & 1.055(76) & 1.07  \\
40 & 0.99(11)  & 1.09  \\
48 & 0.99(16)  & 1.16  \\
54 & 1.02(22)  & 1.23  \\
64 & 0.76(29)  & 1.14  \\
80 & 0.25(33)  & 0.76  \\
96 & 0.06(38)  & 0.74  \\
108 & 0.27(46) & 0.70  \\
128 & 0.11(59) & 0.75  \\
\hline
\end{tabular}
\end{center}
\end{table}

We have fit the quantity $\Phi_2(L)$ to~(\ref{eq:phi}) within $L \in [L_{\mathrm{min}},1024]$ 
(estimated from the $\chi/L^2$ data within $L \in [L_{\mathrm{min}}/2,2048]$) to evaluate $\omega$.
The results are collected in Tab.~\ref{tab2}.
The estimated $\omega$ values are essentially decreased for $L_{\mathrm{min}} \ge 80$
as compared to smaller $L_{\mathrm{min}}$ values. Moreover, the quality of fits is 
remarkably improved in this case, i.~e., 
the values of $\chi^2$ of the fit per degree of freedom ($\chi^2/\mathrm{d.o.f.}$) become smaller.
Note that $L_{\mathrm{min}} = 80$ corresponds to the fit interval for $\Phi_2(L)$ in Fig.~\ref{fig1},
where the data are well consistent with $\omega=1/8$.
From a formal point of view, $\omega = 0.25(33)$ at $L_{\mathrm{min}} = 80$ can be considered as the 
best estimate from our data, since it perfectly agrees with the results for $L_{\mathrm{min}} > 80$ and has the minimal
statistical error within $L_{\mathrm{min}} \ge 80$. The estimate $\omega = 0.06(38)$ at $L_{\mathrm{min}} =96$
most clearly shows the deviation below the usually accepted values at about $0.8$,
e.~g., $\omega = 0.832(6)$ reported in~\cite{Has1}.
Our estimation is fully consistent with the analytical arguments in Sec.~\ref{sec:analytical},
since all our $\omega$ values for $L_{\mathrm{min}} \ge 80$ are smaller than $\omega_{\mathrm{max}} \approx 0.38$
and also well agree with $1/8$.

Unfortunately, the statistical accuracy of this estimation is too low to rule out a possibility that the
dropping of $\omega$ to smaller values at $L_{\mathrm{min}} \ge 80$ is caused by statistical errors
in the data. 
However, the decrease in $\omega$ for large enough lattice sizes is strongly supported
by the theorem discussed in Sec.~\ref{sec:analytical}. Note also that the recent MC analysis of 
the 2D $\varphi^4$ model~\cite{KMR_14} is consistent with this theorem. These facts
make our MC estimation plausible. 

Note that there exist many quantities, which scale asymptotically as $A + B L^{-\omega}$ with
different values of coefficients $A$ and $B$ --- see, e.~g., \cite{Hasenbusch,Has1}, as well as
the examples in the next section. In principle, all of them can be used to estimate $\omega$.
However, it is possible that the leading correction term $B L^{-\omega}$ for a subset of
such quantities is too small as compared to statistical errors and, therefore, it is not well detectable at large lattice sizes. 
It means that a correction with small $\omega$ of about $1/8$, probably, will not be detected by MC analysis
in many cases, but this still does not imply that such a correction does not exist.
Thus, it is sufficient to demonstrate clearly that such a correction exists in one of the cases.
Our MC analysis shows that $\Phi_2(L)$ is an appropriate quantity where corrections, i.~e.,
variations in $\Phi_2(L)$, are well detectable even for very large values of $L$.
Moreover, it suggests that a correction with such small $\omega$ as $1/8$, very likely,
exists here.

\subsection{Different estimates from the data for smaller lattice sizes}
\label{sec:difomega}

The quality of fits with $L_{\mathrm{min}} < 80$ is remarkably improved
if 5 data points for the largest lattice sizes are discarded, i.~e., if $\Phi_2(L)$ is fit within
$L \in [L_{\mathrm{min}},432]$. Choosing also not too large
values of $L_{\mathrm{min}}$, we obtain formally quite good 
(provided by good fits with sufficiently small $\chi^2/\mathrm{d.o.f.}$ values) 
and stable estimates from 
remarkably smaller lattice sizes than those in Sec.~\ref{sec:large}. These are presented
in Tab.~\ref{tab3}. The estimate $\omega = 1.171(96)$  at $L_{\mathrm{min}}=32$ is accepted as the best one
from this reduced data set, since it perfectly agrees with the results for $L_{\mathrm{min}}>32$ and has the
smallest statistical error.

\begin{table}
\caption{The values of $\omega$, extracted from the fits of $\Phi_2(L)$ 
to~(\ref{eq:phi}) within $L \in [L_{\mathrm{min}},432]$, together with
the values of $\chi^2/\mathrm{d.o.f.}$ of the fits.}
\label{tab3}
\begin{center}
\begin{tabular}{|c|c|c|}
\hline
\rule[-2mm]{0mm}{7mm}
$L_{\mathrm{min}}$ & $\omega$  & $\chi^2/\mathrm{d.o.f.}$ \\
\hline
32 & 1.171(96) & 0.77  \\
40 & 1.17(14)  & 0.83  \\
48 & 1.29(22)  & 0.84  \\
\hline
\end{tabular}
\end{center}
\end{table}

We have tested another finite--size scaling method. Based on our simulations discussed 
in~\cite{KMR_2013}, we have evaluated $U=U(L)$ at the pseudocritical coupling $\hat{\beta}_c(L)$, corresponding
to the maximum of specific heat $C_V$. It scales as
\begin{equation}
 U(L) = {\cal A} + {\cal B} L^{-\omega}
 \label{eq:UL}
\end{equation}
at large $L$. This method is similar in spirit to 
the one used by Hasenbusch for the 3D $\varphi^4$ model in~\cite{Hasenbusch}. 
The only difference is that another pseudocritical
coupling (corresponding to certain value of $Z_a/Z_p$, where $Z_p$ and $Z_a$ are partition
functions for the lattice with periodic and antiperiodic boundary conditions) has been
used in~\cite{Hasenbusch}. We have found that our $U(L)$ data provide a good fit to~(\ref{eq:UL})
within $8 \le L \le 384$, where $L=384$ is similar to the maximal size $L=360$ simulated in~\cite{Has1}.
These data are listed in Tab.~\ref{tab4}, and the fit results are presented in Tab.~\ref{tab5}.
\begin{table}
\caption{The pseudo-critical couplings
$\hat{\beta}_c$ and the values of $U$ at $\beta=\hat{\beta}_c$
depending on the linear system size $L$.}
\label{tab4}
\begin{center}
\begin{tabular}{|c|c|c|}
\hline
\rule[-2mm]{0mm}{7mm}
L & $\hat{\beta}_c$ & $U$  \\
\hline
384  & 0.22167526(52) & 1.1884(62)  \\
320  & 0.22168192(69) & 1.1901(63)  \\
256  & 0.22169312(76) & 1.1937(50)  \\
192  & 0.2217149(10)  & 1.1940(42)   \\
160  & 0.2217347(14)  & 1.1951(44)   \\
128  & 0.2217742(16)  & 1.1831(32)   \\
96   & 0.2218366(24)  & 1.1917(33)   \\
80   & 0.2219002(32)  & 1.1885(32)   \\
64   & 0.2220057(42)  & 1.1888(30)   \\
48   & 0.2221987(58)  & 1.1930(27)   \\
40   & 0.2223761(76)  & 1.1933(26)   \\
32   & 0.222659(10)   & 1.1983(26)   \\
24   & 0.223195(12)   & 1.2035(19)   \\
20   & 0.223686(13)   & 1.2051(16)  \\
16   & 0.224443(15)   & 1.2121(13)  \\
12   & 0.225813(16)   & 1.22147(93)  \\
10   & 0.226903(18)   & 1.23159(86)  \\
8    & 0.228567(20)   & 1.24474(64)  \\
\hline
\end{tabular}
\end{center}
\end{table}
\begin{table}
\caption{The values of $\omega$, extracted from the fits of $U(L)$ 
to~(\ref{eq:UL}) within $L \in [L_{\mathrm{min}},384]$, together with
the values of $\chi^2/\mathrm{d.o.f.}$ of the fits.}
\label{tab5}
\begin{center}
\begin{tabular}{|c|c|c|}
\hline
\rule[-2mm]{0mm}{7mm}
$L_{\mathrm{min}}$ & $\omega$  & $\chi^2/\mathrm{d.o.f.}$ \\
\hline
8  & 1.247(73) & 0.87  \\
10 & 1.31(12)  & 0.90  \\
12 & 1.24(16)  & 0.94  \\
16 & 1.46(29)  & 0.95  \\
\hline
\end{tabular}
\end{center}
\end{table}
The estimate $\omega=1.247(73)$ at $L_{\mathrm{min}}=8$ seems to be the best one,
as it has the smallest statistical error, a good fit quality,
and it perfectly agrees with the results for $L_{\mathrm{min}}>8$.
This value disagrees (the discrepancy is $5.7$ standard deviations) with the best estimate $\omega = 0.832(6)$ 
of~\cite{Has1}, obtained by a different finite--size scaling method. It well agrees with the 
other value $\omega = 1.171(96)$ reported here.

Searching for a different method, we have evaluated the Fisher zeros of partition function
from MC simulations by the Wolff single cluster algorithm, following the method described in~\cite{GKR11}. The results
for $4 \le L \le 72$ have been reported in~\cite{GKR11}. We have performed high statistics simulations 
(with MC measurements after each $\max\{2,L/4\}$ Wolff clusters, omitting $10^6$ measurements from the 
beginning of each simulation run, and 
totally $5 \times 10^8$ measurements used in the analysis for each $L$) for $4 \le L \le 128$.
Two different pseudo-random number generators, discussed and tested in~\cite{KMR_2011}, have been used to
verify that the results agree within error bars of about one or, sometimes, two standard deviations.
Considering $\beta = \eta + i \xi$ as a complex number, the results for the first Fisher zero 
$\mathrm{Re} \, u^{(1)} + i \, \mathrm{Im} \, u^{(1)}$ in terms of $u=\exp(-4 \beta)$
are reported in Tab.~\ref{tabzero1}. Our values are obtained, evaluating
$R = \langle \cos(\xi E) \rangle_{\eta} + i \langle \sin(\xi E) \rangle_{\eta}$   
(where $E$ is energy) by the histogram reweighting method and minimizing $\mid R \mid$ 
(see~\cite{GKR11}). Reliable results are ensured
by the fact that, for each $L$, the simulation is performed at the coupling $\beta_{\mathrm{sim}}$ which  
is close to  $\mathrm{Re} \beta^{(1)}$ -- see Tab.~\ref{tabzero1}. We have reached it by using the results 
of~\cite{GKR11} and finite--size extrapolations. 
\begin{table}
\caption{The real and imaginary parts of the first Fisher zeros for $u=\exp(-4 \beta)$ (columns 4--5) vs 
lattice size $L$, evaluated from simulations at $\beta=\beta_{\mathrm{sim}} \approx \mathrm{Re} \beta^{(1)}$ 
(columns 2--3).} 
\label{tabzero1}
\begin{center}
\begin{tabular}{|c|c|c|c|c|}
\hline
\rule[-2mm]{0mm}{7mm}
$L$ &  $\beta_{\mathrm{sim}}$ & $\mathrm{Re} \beta^{(1)}$ &
$\mathrm{Re} \, u^{(1)}$ & $\mathrm{Im} \, u^{(1)}$
\\
\hline
4  & 0.2327517   & 0.2327392(37) & 0.3842870(59) & -0.0877415(55) \\
6  & 0.228982187 & 0.2289856(28) & 0.3975550(44) & -0.0454038(44) \\
8  & 0.22674832  & 0.2267531(27) & 0.4027150(44) & -0.0285905(42) \\
12 & 0.224558048 & 0.2245557(17) & 0.4070191(28) & -0.0149314(25) \\
16 & 0.223560276 & 0.2235605(12) & 0.4088085(19) & -0.0094349(17) \\
24 & 0.22268819  & 0.22268780(72) & 0.4103176(12) & -0.0049422(11) \\
32 & 0.222317896 & 0.22231846(49) & 0.41094218(81) & -0.00312478(84) \\
48 & 0.22200815  & 0.22200835(26) & 0.41146087(43) & -0.00163982(49) \\
64 & 0.221880569 & 0.22188039(17) & 0.41167349(29) & -0.00103825(37) \\
96 & 0.2217737   & 0.22177375(12) & 0.41185008(21) & -0.00054521(19) \\
128 & 0.22173025 & 0.221730228(83) & 0.41192200(14) & -0.00034552(14) \\
\hline
\end{tabular}
\end{center}
\end{table}
We have also estimated the second zeros for $L=4, 32, 64$ from different simulation runs
-- see Tab.~\ref{tabzero2}.
\begin{table}
\caption{The real and imaginary parts of the second Fisher zeros for $u=\exp(-4 \beta)$ (columns 4--5) vs 
lattice size $L$, evaluated from simulations at $\beta=\beta_{\mathrm{sim}} \approx \mathrm{Re} \beta^{(2)}$ 
(columns 2--3).}
\label{tabzero2}
\begin{center}
\begin{tabular}{|c|c|c|c|c|}
\hline
\rule[-2mm]{0mm}{7mm}
$L$ &  $\beta_{\mathrm{sim}}$ & $\mathrm{Re} \beta^{(2)}$ &
$\mathrm{Re} \, u^{(2)}$ & $\mathrm{Im} \, u^{(2)}$
\\
\hline
4  & 0.2464072   & 0.246484(18)  & 0.344470(27) & -0.143307(23) \\
32 & 0.22313686  & 0.223169(15)  & 0.409529(25) & -0.004891(25) \\
64 & 0.222166355 & 0.2221781(85) & 0.411182(14) & -0.001616(13) \\
\hline
\end{tabular}
\end{center}
\end{table}

Our results in Tabs.~\ref{tabzero1} and~\ref{tabzero2} are reasonably
consistent with those of~\cite{GKR11}, but are more accurate and include larger lattice sizes.
Like in~\cite{GKR11}, the results for the second zeros are much less accurate than those for
the first zeros. Therefore only the latter ones are used here in the analysis, considering
the ratios $\Psi_1(L) = \mathrm{Im} \, u^{(1)}(L) / (\mathrm{Re} \, u^{(1)}(L) - u_c )$ and 
$\Psi_2(L) = \mathrm{Im} \, u^{(1)}(L) / \mathrm{Im} \, u^{(1)}(L/2)$, which behave asymptotically
as $A + B L^{-\omega}$ at $L \to \infty$. Here $u_c = \exp(-4 \beta_c)$ is the critical $u$ value,
corresponding to the critical coupling $\beta_c$. The estimation of correction--to--scaling
exponent $\omega$ from fits of $\Psi_1(L)$ to  $A + B L^{-\omega}$ has been considered in~\cite{GKR11},
assuming the known approximate value $0.2216546$ of $\beta_c$. The use of $\Psi_2(L)$ instead
of $\Psi_1(L)$ is another method, which has an advantage that it does not require the knowledge
of the critical coupling $\beta_c$. However, a disadvantage is that the data for two sizes, $L$ and $L/2$,
are necessary for one value of $\Psi_2(L)$. The values of $\Psi_1(L)$ and $\Psi_2(L)$ are listed
in Tab.~\ref{tabPsi}. The standard errors of $\Psi_1(L)$ are calculated by the jackknife 
method~\cite{MC}, 
thus taking into account the statistical correlations between $\mathrm{Re} \, u^{(1)}$ and 
$\mathrm{Im} \, u^{(1)}$. As in~\cite{GKR11}, the errors due to the uncertainty in $\beta_c$ are ignored, assuming 
that $\beta_c = 0.2216546$ holds with a high enough accuracy. 
According to~\cite{KMR_2011}, this $\beta_c$ value, likely, is correct within error bars of about $\pm 3 \times 10^{-8}$.
It justifies the actual estimation.
\begin{table}
\caption{The ratios $\Psi_1(L) = \mathrm{Im} \, u^{(1)}(L) / (\mathrm{Re} \, u^{(1)}(L) - u_c )$ and 
$\Psi_2(L) = \mathrm{Im} \, u^{(1)}(L) / \mathrm{Im} \, u^{(1)}(L/2)$ depending on the lattice size $L$.}
\label{tabPsi}
\begin{center}
\begin{tabular}{|c|c|c|}
\hline
\rule[-2mm]{0mm}{7mm}
$L$ & $\Psi_1$  & $\Psi_2$ \\
\hline
8  & 3.0638(15)  & 0.325829(52)  \\
12 & 2.9698(17)  & 0.328858(64)  \\
16 & 2.9136(18)  & 0.330001(77)  \\
24 & 2.8581(21)  & 0.330994(92)  \\
32 & 2.8289(22)  & 0.33119(11)  \\
48 & 2.7988(23)  & 0.33180(12)  \\
64 & 2.7814(24)  & 0.33226(15)  \\
96 & 2.7718(31)  & 0.33248(15)  \\
128 & 2.7691(33) & 0.33279(18)  \\\hline
\end{tabular}
\end{center}
\end{table}

\begin{table}
\caption{The values of $\omega$, extracted from the fits of $\Psi_1(L)$ 
in Tab.~\ref{tabPsi} to $A + B L^{-\omega}$ within $L \in [L_{\mathrm{min}},L_{\mathrm{max}}]$, together with
the values of $\chi^2/\mathrm{d.o.f.}$ of the fits.}
\label{tabPsi1fit}
\begin{center}
\begin{tabular}{|c|c|c|c|}
\hline
\rule[-2mm]{0mm}{7mm}
$L_{\mathrm{max}}$ & $L_{\mathrm{min}}$ & $\omega$  & $\chi^2/\mathrm{d.o.f.}$ \\
\hline
    & 8  & 0.807(27) & 1.01  \\
64  & 12 & 0.903(60) & 0.22  \\
    & 16 & 0.84(10)  & 0.06  \\ 
 \hline
    & 8  & 0.872(21)  & 3.12  \\
128 & 12 & 0.997(42)  & 1.21  \\
    & 16 & 1.026(65)  & 1.43  \\
\hline
\end{tabular}
\end{center}
\end{table}
The values of exponent $\omega$, extracted from the fits of $\Psi_1(L)$ to $A + B L^{-\omega}$ within
$L \in [L_{\mathrm{min}},L_{\mathrm{max}}]$, are collected in Tab.~\ref{tabPsi1fit}. The results $\omega=0.903(60)$ 
for $L \in [12,64]$ and $\omega=0.84(10)$ for $L \in [16,64]$ agree within error bars with the results for 
similar fit intervals in~\cite{GKR11}, i.~e., $\omega = 0.77(9)$ for 
$L \in [12,72]$ and $\omega = 0.63(16)$ for $L \in [16,72]$. However, the fits with
$L_{\mathrm{max}}=128$ are preferable for a reasonable estimation of the asymptotic exponent $\omega$. 
The best estimate with $L_{\mathrm{max}}=128$ is $\omega = 0.997(42)$, obtained at
$L_{\mathrm{min}}=12$. Indeed, this fit has an acceptable
$\chi^2/\mathrm{d.o.f.}$ value and the result is well consistent with that for $L_{\mathrm{min}}=16$,
where the statistical error is larger. It turns out that the estimated value of $\omega$ becomes larger when 
$L_{\mathrm{max}}$ is increased from $64$ to $128$. 
One of possible explanations, which is consistent with
the data in Tab.~\ref{tabPsi}, is such that
the $\Psi_1(L)$ plot has a minimum near $L=128$ or at somewhat larger $L$ values. In this case
the actual method is really valid only for remarkably larger lattice sizes.

The results of the other method, using the ratio $\Psi_2(L)$ instead of $\Psi_1(L)$,
are collected in Tab.~\ref{tabPsi2fit}. Here $L_{\mathrm{max}}=128$ is fixed and only $L_{\mathrm{min}}$ is varied.
The standard errors of $\omega$  are calculated, taking into account that fluctuations in $\mathrm{Im} \, u^{(1)}$
are the statistically independent quantities. As we can see, the estimated exponent $\omega$ decreases with
increasing of $L_{\mathrm{min}}$ in the considered range. Since $\chi^2/\mathrm{d.o.f.}$ is about unity
for moderately good fits, the estimates $\omega=0.61(19)$ at $L_{\mathrm{min}}=16$ and $\omega=0.18(37)$ 
at $L_{\mathrm{min}}=24$ are acceptable.   
\begin{table}
\caption{The values of $\omega$, extracted from the fits of $\Psi_2(L)$ 
in Tab.~\ref{tabPsi} to $A + B L^{-\omega}$ within $L \in [L_{\mathrm{min}},128]$, together with
the values of $\chi^2/\mathrm{d.o.f.}$ of the fits.}
\label{tabPsi2fit}
\begin{center}
\begin{tabular}{|c|c|c|}
\hline
\rule[-2mm]{0mm}{7mm}
$L_{\mathrm{min}}$ & $\omega$  & $\chi^2/\mathrm{d.o.f.}$ \\
\hline
 8   & 1.400(57) & 4.12  \\
 12  & 0.96(13)  & 2.02  \\
 16  & 0.61(19)  & 1.20  \\ 
 24  & 0.18(37)  & 0.88  \\
\hline
\end{tabular}
\end{center}
\end{table}

Summarizing the results of this section, we conclude that three
of the considered here methods give larger values of $\omega$ 
($1.171(96)$, $1.274(72)$ and $0.997(42)$) than  $\omega=0.832(6)$
reported in~\cite{Has1}, whereas the fourth method
tends to give smaller values ($0.61(19)$ and $0.18(37)$).
Thus, it is evident that the estimation of $\omega$ from finite--size scaling, using
the data for not too large lattice sizes (comparable with $L \le 360$ in~\cite{Has1}
or $L \le 72$ in~\cite{GKR11}) does not lead to conclusive results.
Indeed, the obtained values depend on the particular method chosen and
are varied from $1.274(72)$ to $0.18(37)$ in our examples.

\section{Influence of $\omega$ on the estimation of exponents $\eta$ and $\nu$}
\label{sec:influence}

Allowing a possibility that the correction--to--scaling
exponent $\omega$ of the 3D Ising model is, indeed, essentially smaller than the commonly accepted values of about $0.8$,  
we have tested the influence of $\omega$ on the estimation of critical exponents $\eta$ and $\nu$
(or $1/\nu$). We have fit our susceptibility data at $\beta = \widetilde{\beta}_c(L)$ to the ansatz
\begin{equation}
 \chi = L^{2-\eta} \left( a_0 + \sum\limits_{k=1}^m a_k L^{-k \omega} \right)
\label{eq:chiansatz}
\end{equation}
with $m=1$ and $m=2$ to estimate $\eta$ at three fixed values of the exponent $\omega$, i.~e.,
$\omega=0.8$, $\omega=0.38$ and $\omega=1/8$. The first one is very close to the known RG value 
$\omega=0.799 \pm 0.011$~\cite{Justin} and also is quite similar to a more recent RG estimate
$\omega=0.782(5)$~\cite{PS08} and the MC estimate $\omega=0.832(6)$ of~\cite{Has1}. The second value corresponds to the upper bound 
$\omega_{\mathrm{max}} \approx 0.38$ stated in Sec.~\ref{sec:analytical}, and the third value
$1/8$ is extracted from the GFD theory~\cite{K_Ann01,K2012}. 
There exist different corrections to scaling, but the two correction terms in~(\ref{eq:chiansatz}) are the most
relevant ones at $L \to \infty$, as it can be seen from the analysis in~\cite{KMR_2013,Has1}.
The results of the fit within $L \in [L_{\mathrm{min}},2048]$ depending on $\omega$, $m$ and $L_{\mathrm{min}}$
are shown in Tab.~\ref{tab6}.
Similarly, we have fit our $\partial Q/\partial \beta$ data at $\beta = \widetilde{\beta}_c(L)$ to the ansatz
\begin{equation}
 \frac{\partial Q}{\partial \beta} = L^{1/\nu} \left( b_0 + \sum\limits_{k=1}^m b_k L^{-k \omega} \right)
\label{eq:Qansatz}
\end{equation}
and have presented the results in Tab.~\ref{tab7}.

\begin{table}
\caption{The critical exponent $\eta$, extracted from the fit of the susceptibility data 
at $\beta = \widetilde{\beta}_c(L)$ to the ansatz~(\ref{eq:chiansatz})
within $L \in [L_{\mathrm{min}},2048]$, as well as the $\chi^2/\mathrm{d.o.f.}$ of the fit 
depending on $\omega$, $m$ and $L_{\mathrm{min}}$.}
\label{tab6}
\begin{center}
\begin{tabular}{|c|c|c|c|c|}
\hline
\rule[-2mm]{0mm}{7mm}
$\omega$ & $m$ & $L_{\mathrm{min}}$ & $\eta$  & $\chi^2/\mathrm{d.o.f.}$ \\
\hline
       &   & 32 & 0.03617(45)  & 0.93  \\
       & 1 & 48 & 0.03562(59)  & 0.89  \\
 0.8   &   & 64 & 0.03563(76)  & 0.97  \\ 
\cline{2-5}
       &   & 32 & 0.03521(94) & 0.91  \\
       & 2 & 48 & 0.0366(14)  & 0.90  \\
       &   & 64 & 0.0384(18)  & 0.86  \\ 
\hline
       &   & 32 & 0.04387(78) & 1.40  \\
       & 1 & 48 & 0.0414(11)  & 0.82  \\
0.38   &   & 64 & 0.0407(14)  & 0.84  \\ 
\cline{2-5}
       &   & 32 & 0.0342(32)  & 0.97  \\
       & 2 & 48 & 0.0408(45)  & 0.86  \\
       &   & 64 & 0.0465(60)  & 0.84  \\ 
\hline
       &   & 32 & 0.0656(15)  & 1.90  \\
       & 1 & 48 & 0.0589(22)  & 0.85  \\
$1/8$  &   & 64 & 0.0562(30)  & 0.80  \\ 
\cline{2-5}
       &   & 32 & 0.0106(16)  & 1.51  \\
       & 2 & 48 & 0.031(54)   & 0.87  \\
       &   & 64 & 0.075(30)   & 0.83  \\ 
\hline
\end{tabular}
\end{center}
\end{table}

\begin{table}
\caption{The critical exponent $1/\nu$, extracted from the fit of the $\partial Q/\partial \beta$ data 
at $\beta = \widetilde{\beta}_c(L)$ to the ansatz~(\ref{eq:Qansatz})
within $L \in [L_{\mathrm{min}},2048]$, as well as the $\chi^2/\mathrm{d.o.f.}$ of the fit  
depending on $\omega$, $m$ and $L_{\mathrm{min}}$.}
\label{tab7}
\begin{center}
\begin{tabular}{|c|c|c|c|c|}
\hline
\rule[-2mm]{0mm}{7mm}
$\omega$ & $m$ & $L_{\mathrm{min}}$ & $1/\nu$  & $\chi^2/\mathrm{d.o.f.}$ \\
\hline
    &   & 32 & 1.5872(16)  & 0.63  \\
    & 1 & 48 & 1.5895(22)  & 0.48  \\
0.8 &   & 64 & 1.5880(27)  & 0.47  \\ 
\cline{2-5}
    &   & 32 & 1.5914(34)  & 0.56  \\
    & 2 & 48 & 1.5854(49)  & 0.46  \\
    &   & 64 & 1.5869(64)  & 0.50  \\ 
\hline
     &   & 32 & 1.5873(29)  & 0.63  \\
     & 1 & 48 & 1.5913(40)  & 0.49  \\
0.38 &   & 64 & 1.5880(52)  & 0.47  \\ 
\cline{2-5}
     &   & 32 & 1.598(12)  & 0.61  \\
     & 2 & 48 & 1.576(15)  & 0.47  \\
     &   & 64 & 1.584(22)  & 0.50  \\ 
\hline
     &   & 32 & 1.5878(84)  & 0.63  \\
     & 1 & 48 & 1.599(14)   & 0.50  \\ 
$1/8$ &  & 64 & 1.588(15)   & 0.47  \\ 
\cline{2-5}
     &   & 32 & 1.636(21)  & 0.64  \\
     & 2 & 48 & 1.525(50)  & 0.47  \\
     &   & 64 & 1.56(37)   & 0.50  \\ 
\hline
\end{tabular}
\end{center}
\end{table}

Considering the fits with only the leading correction to scaling included ($m=1$),
one can conclude from Tab.~\ref{tab6} that the estimated critical exponent $\eta$ 
increases with decreasing of $\omega$, whereas the exponent $1/\nu$ in
Tab.~\ref{tab7} is rather stable. The sub-leading correction to
scaling ($m=2$) makes the estimated exponents $\eta$ and $1/\nu$ remarkably less stable
for small $\omega$ values, such as $\omega=1/8$. The latter value is expected from the GFD 
theory~\cite{K_Ann01,K2012}, so that the estimation at $\omega=1/8$ is self-consistent
within this approach. In this case, the estimation 
of $\eta$ appears to be compatible with the theoretical value $\eta=1/8=0.125$ of~\cite{K_Ann01},
taking into account that the evaluated $\eta$ increases with $L_{\mathrm{min}}$.
Moreover, the self-consistent estimation of $1/\nu$ is even very well consistent with $\nu = 2/3$ predicted in~\cite{K_Ann01}.
In particular, $1/\nu = 1.525(50)$ can be considered as the best estimate at $\omega=1/8$ and $m=2$
(it has the smallest  $\chi^2/\mathrm{d.o.f.}$ value and much smaller statistical error
than the estimate at $L_{\mathrm{min}}=64$), which well agrees with $1.5$.
Thus, contrary to the statements in~\cite{GKR11}, the value $\nu=2/3$ of the GFD theory is not ruled out,
since it is possible that $\omega$ has a much smaller value than $0.832(6)$ assumed in~\cite{GKR11}.

A question can arise about the influence of $\omega$ value on the estimation of critical exponents 
in the case of improved Hamiltonians~\cite{Has1,Has2}. It is expected that the leading corrections to scaling vanish in this case,
and therefore the influence of $\omega$ is small. However, in the case if the asymptotic corrections to scaling
are described by the exponent $\omega \le \omega_{\mathrm{max}} \approx 0.38$, as it is strongly suggested by
the theorem discussed in Sec.~\ref{sec:analytical}, the vanishing of leading corrections cannot be supported by the existing
MC analyses of such models. Indeed, in these analyses the asymptotic corrections to scaling are not correctly identified
(probably, because of too small lattice sizes) if $\omega \le \omega_{\mathrm{max}} \approx 0.38$, since one finds that $\omega \approx 0.8$.

\section{Comparison of recent results}
\label{sec:compare}

It is interesting to compare our MC estimates and those of~\cite{Has1}
with the most recent RG (3D expansion) values of~\cite{PS08} cited in~\cite{Has1}.
Note that the estimates of $\omega$ in~\cite{Has1} and~\cite{PS08}, i.~e.,
$\omega=0.832(6)$ and $\omega=0.782(5)$, are clearly inconsistent within the claimed error bars.
This discrepancy, however, can be understood from the point of view of our MC analysis,
suggesting that the real uncertainty in the MC estimation of $\omega$ can be rather large.

\begin{table}
\caption{Recent estimates of the critical exponents $\eta$ and $\nu$ from different sources.
Our values correspond to $\omega=0.8$, $m=2$ and $L_{\mathrm{min}}=64$.}
\label{tab8}
\begin{center}
\begin{tabular}{|c|c|c|c|}
\hline
source & method & $\eta$  & $\nu$ \\
\hline
this work        & MC & 0.0384(18)  & 0.6302(25)  \\
Ref.~\cite{Has1} & MC & 0.03627(10) & 0.63002(10)  \\
Ref.~\cite{PS08} & 3D exp & 0.0318(3)   & 0.6306(5)  \\
\hline
\end{tabular}
\end{center}
\end{table}

The comparison of critical exponents $\eta$ and $\nu$ is provided in Tab.~\ref{tab8}.
This comparison includes only some recent or relatively new results, since older ones
are extensively discussed in~\cite{Has1,Has2,HasRev}.
Our values correspond to the fits within $L \in [64,2048]$ at $m=2$ and $\omega=0.8$.
The choice of $\omega=0.8$ is reasonable here, since this value is close enough to
the above mentioned estimates $\omega=0.832(6)$ and $\omega=0.782(5)$, and practically the same
results are obtained if $\omega=0.8$ is replaced by any of these two values.
According to the claimed statistical error bars, the estimates of~\cite{Has1} seem to be extremely accurate.
Note, however, that these estimates are extracted from much smaller lattice sizes ($L \le 360$) as compared 
to ours ($L \le 2048$). 

The values of $\nu$ in Tab.~\ref{tab8} are consistent with each other.
The MC estimates of $\eta$ are consistent, as well. However, the recent RG value of~\cite{PS08}
appears to be somewhat smaller and not consistent within the error bars with the actual MC estimations,
even if the assumed values of $\omega$ are about $0.8$, as predicted by the perturbative RG theory.
In particular, the discrepancy with the MC value of~\cite{Has1} is about $45$ standard 
deviations of the MC estimation or about $15$ error bars of the RG estimation.

Recently, the conformald field theory (CFT) has been applied to the 3D Ising model~\cite{Showk} 
to obtain very accurate values of the critical exponents, using the numerical conformal bootstrap method. 
The conformal--symmetry relations for the correlation functions, like (2.1) in~\cite{Showk}, are known to hold asymptotically  
in two dimensions, whereas their validity in 3D case can be questioned. 
Here ``asymptotically'' means that the limit $L/x \to \infty$, $\xi/x \to \infty$ and $a/x \to 0$
is considered, where $x$ is the actual distance, $a$ is the lattice spacing and $\xi$ is the correlation length.
In other words, the conformal symmetry is expected to hold exactly for the asymptotic correlation functions
on an infinite lattice ($L = \infty$) at the critical point ($\beta = \beta_c$). These 
asymptotic correlation functions are obtained by subtracting from the exact correlations functions (at $L=\infty$
and $\beta=\beta_c$) the corrections to scaling,
containing powers of $a/x$.
The existence of the 
conformal symmetry in the 3D Ising model has been supported by a non--trivial MC test in~\cite{MCconf}.

Apart from the assumption of the validity of (2.1) in~\cite{Showk} for the 3D Ising model, 
the following hypotheses have been proposed: 
\begin{enumerate}
\item[(i)] 
There exists a sharp kink on the border of the two--dimensional region of the allowed values of 
the operator dimensions $\Delta_{\sigma} = (1+\eta)/2$ and $\Delta_{\epsilon} = 3 - 1/\nu$; 
\item[(ii)] 
Critical exponents of the 3D Ising model correspond just to this kink.
\end{enumerate}

These hypotheses have been supported by the MC estimates 
of the exponents $\eta$, $\nu$ and $\omega$ in~\cite{Has1}. However, the obtained in~\cite{Showk} exponent $\omega=0.8303(18)$ 
is not supported by our MC value $\omega = 0.25(33)$, obtained from the susceptibility data for very large lattice sizes $L \le 2048$. 
Moreover, it does not satisfy the inequality $\omega \le (\gamma-1)/\nu$, following from the theorem discussed in Sec.~\ref{sec:analytical}. 
This apparent contradiction can be understood from the point of view that corrections to scaling are not fully controlled
in the CFT. 
Indeed, the prediction for $\omega$ in this CFT 
is based on the assumption that $\omega = \Delta_{\epsilon'} -3$ holds, where $\Delta_{\epsilon'}$ is the dimension of an
irrelevant operator in the conformal analysis of the asymptotic four--point correlation function. 
It means that corrections to scaling of the exact four--point correlation function are discarded
(to obtain the asymptotic correlation function, as discussed before) and not included into the analysis.  

More recently, a modified conformal bootstrap analysis has been performed in~\cite{xx}, where the mentioned two hypotheses 
have been replaced with the hypothesis that the operator product expansion (OPE) contains only two relevant scalar operators.
The results for the exponents $\Delta_{\sigma}$ and $\Delta_{\epsilon}$ (or $\eta$ and $\nu$) are consistent with those 
of~\cite{Showk}. This consistency is not surprising, since both methods agree with the idea that the operator spectrum
of the 3D Ising model is relatively simple, so that the true values of  $\Delta_{\sigma}$ and $\Delta_{\epsilon}$ 
are located inside of a certain narrow region (as in~\cite{xx}) or on its border (as in~\cite{Showk}), where many operators are
decoupled from the spectrum. Apparently, the analysis in~\cite{xx} does not lead to a contradiction 
with the two relations $\omega \le (\gamma-1)/\nu$ (the theorem) and  $\omega = \Delta_{\epsilon'} -3$, 
since only $\Delta_{\epsilon'}>3$ is assumed for the dimension  $\Delta_{\epsilon'}$.
Thus, both relations can be satisfied simultaneously, if the 3D Ising point in Fig.~1 of~\cite{Showk}
is located inside of the allowed region, rather than on its border.
This possibility is supported by the behavior of the effective 
exponent $\eta_{\mathrm{eff}}$ in our Fig.~\ref{fig1}. It suggests that the asymptotic exponent $\eta$ could
be larger than it is usually expected from MC simulations for relatively small lattice sizes, as in~\cite{Has1}.
Thus, it is important to make further refined estimations, based on MC data for very large lattice sizes,
in order to verify the hypotheses proposed in~\cite{Showk,xx}.

If the hypothesis~(i) about the existence of a sharp kink is true, then 
this kink, probably, has a special meaning for the 3D Ising model. Its existence, however, is not evident. 
\begin{figure}
\begin{center}
\includegraphics[width=0.6\textwidth]{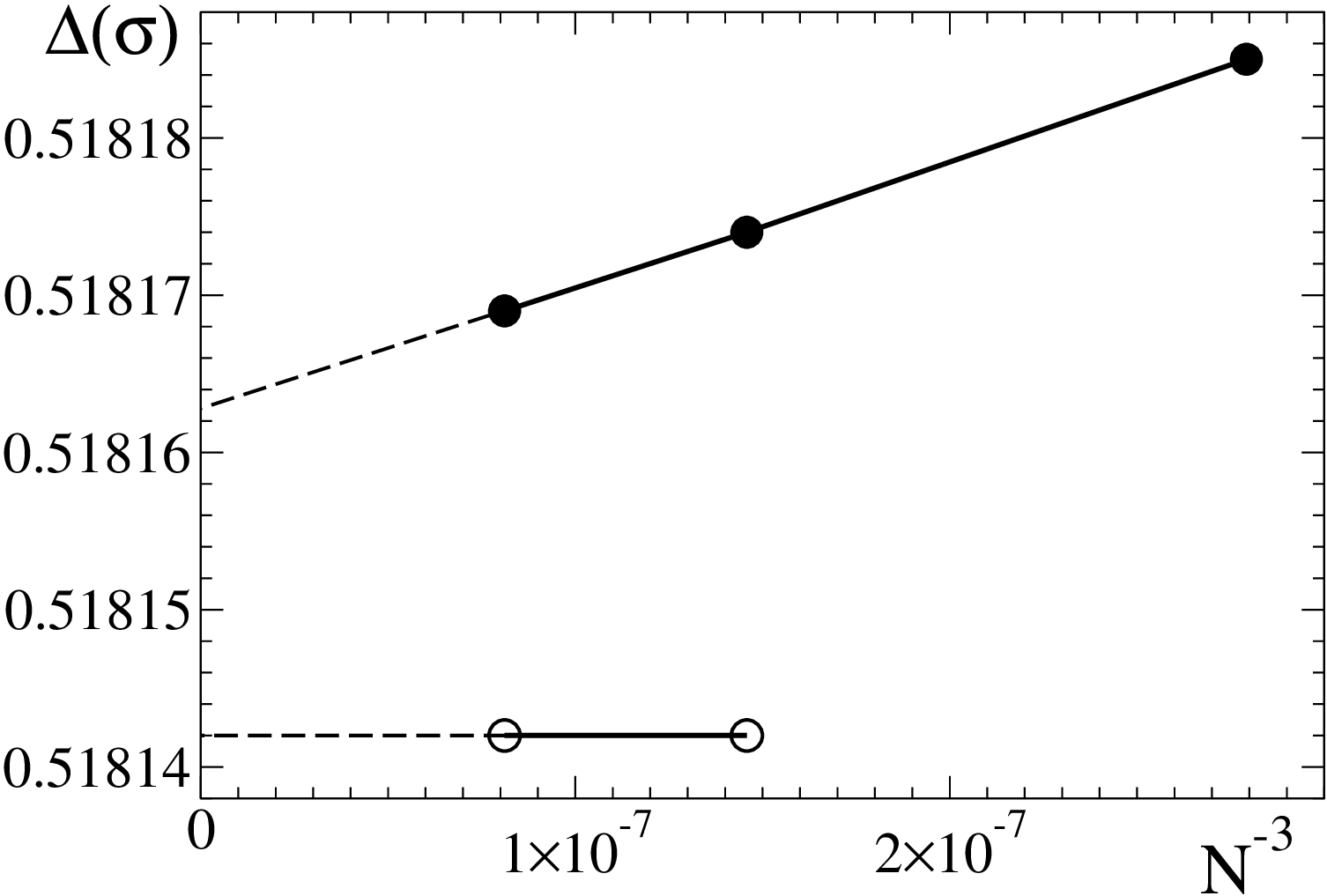}
\end{center}
\caption{The values of $\Delta(\sigma)$, corresponding to the
minimum (solid circles) and the ``kink'' (empty circles) in the plots of Fig.~7 in~\cite{Showk}
depending on $N^{-3}$. The dashed lines show linear extrapolations.}
\label{fig2}
\end{figure}
According to the conjectures of~\cite{Showk}, such a kink is formed at $N \to \infty$, where $N$ is the number of derivatives
included into the analysis. As discussed in~\cite{Showk}, it implies that the minimum in the plots of Fig.~7 
in~\cite{Showk} should be merged with the apparent ``kink''
at $N \to \infty$. This ``kink'' is not really sharp at a finite $N$. 
Nevertheless, its location can be identified with the value of $\Delta(\sigma)$, at which
the second derivative of the plot has a local maximum. The minimum 
of the plot is slightly varied with $N$, whereas the ``kink'' is barely moving~\cite{Showk}.
Apparently, the convergence to a certain asymptotic curve is remarkably faster than $1/N$, as it can be expected 
from Fig.~7  and other similar figures in~\cite{Showk}. In particular, we have found
that the location of the minimum in Fig.~7 of~\cite{Showk} is varied almost linearly with $N^{-3}$.
We have shown it in Fig.~\ref{fig2} by solid circles, the position
of the ``kink'' being indicated by empty circles. The error bars of 
$\pm 0.000001$ correspond to the symbol size. The results for $N=153, 190, 231$ are presented,
skipping the estimate for the location of the ``kink'' at $N=153$, which cannot be well determined
from the corresponding plot in Fig.~7 of~\cite{Showk}.
The linear extrapolations (dashed lines) suggest that the 
minimum, very likely, is moved only slightly closer to the ``kink'' when $N$ is varied from
$N=231$ to $N = \infty$. The linear extrapolation might be too inaccurate. Only in this case  
a refined numerical analysis for larger $N$ values 
can possibly confirm the hypothesis about the formation of a sharp kink at $N \to \infty$.

The results of both~\cite{Showk} and~\cite{xx} strongly support the commonly accepted
3D Ising values of the critical exponents $\eta$ and $\nu$. In particular, 
the estimates $\eta=0.03631(3)$ and $\nu = 0.62999(5)$ have been reported in~\cite{Showk}.
However, these estimates are obtained, based on certain hypotheses. If these
hypotheses are not used, then the conformal bootstrap analysis appears to be consistent even
with the discussed here GFD values $\eta=1/8$ and $\nu =2/3$. Indeed,
the corresponding operator dimensions $\Delta_{\sigma} = (1+\eta)/2$ and $\Delta_{\epsilon} = 3 - 1/\nu$
lie inside of the allowed region in Fig.~1 of~\cite{Showk}.

The hypotheses~(i) and~(ii) can be questioned in view of the observations
summarized in Fig.~\ref{fig2}. The hypothesis of~\cite{xx} about the existence of just two
relevant scalar operators might be supported by some physically--intuitive arguments.
In particular, one needs to adjust two scalar parameters $P$ (pressure) and $T$ (temperature) to reach the critical 
point of a liquid--vapor system. 
A real support for this hypothesis is provided by the already known
estimations of the critical exponents. Taking into account the non--perturbative nature of 
the critical phenomena, the most reliable estimates are based on non--perturbative
methods, such as the Monte Carlo simulation. An essential point in this discussion is that the MC estimates 
can be remarkably changed, if unusually large lattices are considered, as it is shown in our current study.

\section{Summary and conclusions}

Analytical as well as Monte Carlo arguments are provided in this paper, showing that 
corrections to scaling in the 3D Ising model are described by a remarkably smaller
exponent $\omega$ than the usually accepted values of about $0.8$.
The analytical arguments in Sec.~\ref{sec:analytical}, which are based
on a rigorous proof of certain theorem, suggest that $\omega \le (\gamma -1)/\nu$ holds,
implying that $\omega$ cannot be larger than $\omega_{\mathrm{max}}=(\gamma -1)/\nu \approx 0.38$
in the 3D Ising model.
The analytical prediction of the GFD theory~\cite{K_Ann01,K2012} is $\omega=1/8$ in this case.
Our MC estimation of $\omega$ from the susceptibility ($\chi$) data of very large lattices (Sec.~\ref{sec:large}) 
is well consistent with these analytical results.  Numerical values, extracted from the $\chi$ data 
within $40 \le L \le 2048$ and $48 \le L \le 2048$ (or $\Phi_2(L)=2^{-4}\chi(2L)/\chi(L/2)$ data within
$80 \le L \le 1024$ and $96 \le L \le 1024$) are $\omega=0.25(33)$ and $\omega = 0.06(38)$, respectively.
Unfortunately, the statistical errors in $\omega$ are rather large.

As discussed in~\cite{KMR_14}, our analytical predictions generally refer to a subset of $n$-vector models, where spin 
is an $n$-component vector with $n=1$ in two dimensions and $n \ge 1$ in three dimensions.
Our recent MC analysis agrees with these predictions  
for the scalar ($n=1$) 2D $\varphi^4$ model~\cite{KMR_14}, where statistical 
errors are small enough. The 3D case with $n=2$ has been tested in~\cite{K2012},
based on accurate experimental data for specific heat in zero gravity conditions very close to the $\lambda$--transition
point in liquid helium. The test in Sec.~4 of~\cite{K2012} reveals some inconsistency of the data with corrections
to scaling proposed by the perturbative RG treatments, indicating that these corrections decay slower,
i.~e., $\theta=\nu \omega$ is smaller than usually expected. This finding is consistent with the theorem 
discussed in Sec.~\ref{sec:analytical}.
The mentioned here facts emphasize the importance of our MC analysis.

Our proposed values of $\omega$ may seem to be incredible in view of a series of known results, 
yielding $\omega$ at about $0.8$ for the 3D Ising model. 
However, it is meaningful to reconsider these results from several aspects. 
\begin{itemize}
 \item 
 First of all, they disagree with non--perturbative arguments in the form of
the rigorously proven theorem, discussed in Sec.~\ref{sec:analytical}. 
\item
This theorem states
that $\omega \le (\gamma-1)/\nu$, whereas the perturbative RG estimates are essentially larger.
In view of the recent analysis in~\cite{K2012x} (see also the discussions in~\cite{K2012}), 
this discrepancy can be understood as a failure of the standard
perturbative RG methods. The actually discussed (Sec.~\ref{sec:compare}) discrepancy between the recent  RG  and MC estimates of the critical
exponent $\eta$ also points to problems in the perturbative approach. 
Moreover, any perturbative method, also the high- and low-temperature series expansions, can fail to give correct results in critical phenomena, 
since it is not the natural domain of validity of the perturbation theory.
 \item 
The previous MC estimations of $\omega$ are based on simulations of lattices not larger than $L \le 360$
in~\cite{Has1}. We have clearly demonstrated in Sec.~\ref{sec:difomega} that the values obtained from
finite--size scaling with such relatively small (as compared to $L \le 2048$ in our study) lattice sizes
depend on the particular method chosen. 
For example, different estimates ranging from $\omega =1.274(72)$ to $\omega=0.18(37)$ are obtained here,
which substantially deviate from the usually reported values between $0.82$ and $0.87$ (see~\cite{Has1,HasRev}).
\item
Although the recent estimate $\omega=0.8303(18)$ of the conformal bootstrap method~\cite{Showk} is inconsistent
with $\omega \le (\gamma-1)/\nu$, the apparent contradiction can be understood and resolved,
as discussed in Sec.~\ref{sec:compare}. 
\end{itemize}

Taking into account the possibility that the correction--to--scaling exponent $\omega$ can be remarkably smaller 
than the usually accepted values at about $0.8$,
we have tested in Sec.~\ref{sec:influence} the influence of $\omega$ on the estimation of critical exponents
$\eta$ and $\nu$. We have concluded that the effect is remarkable if $\omega$ is changed from $0.8$ to a much smaller value, such
as $\omega=1/8$ of the GFD theory. In this case, the error bars strongly increase, and the estimation becomes 
compatible, or even well consistent, with the predictions
of the GFD theory. In particular, the estimate $1/\nu = 1.525(50)$ agrees with the GFD theoretical value $1.5$.

\section*{Acknowledgments}

We are grateful to Slava Rychkov for the useful communications
concerning the conformal bootstrap method.
This work was made possible by the facilities of the
Shared Hierarchical Academic Research Computing Network
(SHARCNET:www.sharcnet.ca). 
The authors acknowledge the use of resources provided by the
Latvian Grid Infrastructure. For more information, please
reference the Latvian Grid website (http://grid.lumii.lv). 
R. M. acknowledges the support from the
NSERC and CRC program.

\end{document}